\documentclass[epjST,numbook]{svjour}
\usepackage{amsmath}
\usepackage{amssymb}
\usepackage{bm}
\usepackage[normalem]{ulem}
\usepackage{bbm}
\usepackage{mathrsfs}
\usepackage{comment}
\usepackage{dsfont}
\usepackage{multirow}
\usepackage{enumerate}
\usepackage{microtype}
\usepackage{graphics}
\usepackage{xcolor}
\definecolor{philipp}{rgb}{1,.4,.3}
\definecolor{michael}{rgb}{0,.8,.5}

\newcommand\NN{{\mathds{N}}}

\newcommand\RR{{\mathds{R}}}
\newcommand\CC{{\mathds{C}}}

\newcommand{\couple}{\mathscr{U}}
\newcommand{\wmcf}{\mathscr{C}}
\newcommand{\ket}[1]{\left \vert #1 \right >}
\newcommand{\bra}[1]{\left < #1 \right \vert}
\newcommand{\matel}[3]{ \displaystyle \left\langle #1 \right \vert #2 \left\vert  #3 \right\rangle }
\newcommand{\expval}[1]{\left\langle #1 \right\rangle}
\newcommand{\innprod}[2]{ \displaystyle \left\langle #1 \vert #2 \right\rangle }
\newcommand{\impart}[1]{\text{Im}\left[ #1 \right]}

\DeclareMathOperator{\imp }{\mathrm{Im}}
\DeclareMathOperator{\rep}{\mathrm{Re}}

\DeclareMathOperator{\Tr}{{Tr}}

\begin{document}
\title{Reducing backaction when measuring temporal correlations in quantum systems} 
\author{Michael Kastner\inst{1,2}\and Philipp Uhrich\inst{1,2}}
\institute{National Institute for Theoretical Physics (NITheP), Stellenbosch 7600, South Africa\and Institute of Theoretical Physics, Department of Physics, University of Stellenbosch, Stellenbosch 7600, South Africa}
\mail{kastner@sun.ac.za}
\abstract{
Dynamic correlations of quantum observables are challenging to measure due to measurement backaction incurred at early times. Recent work [P.~Uhrich {\em et al.}, Phys.\ Rev.\ A, 96:022127 (2017)] has shown that ancilla-based noninvasive measurements are able to reduce this backaction, allowing for dynamic correlations of single-site spin observables to be measured. We generalise this result to correlations of arbitrary spin observables and extend the measurement protocol to simultaneous noninvasive measurements which allow for real and imaginary parts of correlations to be extracted from a single set of measurements. We use positive operator-valued measures to analyse the dynamics generated by the ancilla-based measurements. Using this framework we prove that special observables exist for which measurement backaction is of no concern, so that dynamic correlations of these can be obtained without making use of ancillas.
} 
\maketitle
\section{Introduction}
\label{intro}

The work presented in this article is an extension of the noninvasive measurement protocol (NIMP) presented in \cite{Uhrich_etal}.
This protocol was developed with the goal of providing a general framework that can be used to measure dynamic (or temporal) two-point correlations of observables in quantum spin-lattices.
A number of experimental platforms exist in which lattice spin models can be simulated \cite{Yan_etal13,Schwarzkopf2011,Lahaye_etal09,Britton_etal12,Ozeri2011,Zeiher_etal2016,Bakr_etal2009}, and measurements of static spin correlations have been reported in, for instance, \cite{Zhang_etal2017,Zeiher_etal2016}. Dynamic correlations in quantum systems have been much less explored. The most straightforward way to obtain such correlations would be to measure an observable $O_1$ at an early time $t_1$ and correlate the outcome with that of a measurement of an observable $O_2$ at a later time $t_2$. While the thus obtained quantity is some kind of dynamic correlation function, it is (as shown in App.~A of \cite{Uhrich_etal}) in general not equal to the unitarily-evolved dynamic correlation function $C=\matel{\psi}{O_1(t_1)O_2(t_2)}{\psi}$ that plays a prominent role in many theoretical approaches. The reason for this discrepancy is the measurement backaction that occurs due to the collapse of the wave function when making a measurement at time $t_1$ and influences the outcome of the measurement at $t_2$.

The NIMP avoids this problem by using an ancillary system (ancilla) to perform noninvasive measurements at the early time $t_1$. The basic idea of a noninvasive measurement is to weakly couple an ancilla spin degree of freedom to the target system at $t_1$, and to subsequently measure the state of this ancilla projectively (for an introduction to ancilla based measurements see Refs.~\cite{Jacobs,NielsenChuang,AsherPeres}). This indirect measurement allows one to reduce measurement backaction, but the price to pay is that only a small amount of information about the observable $O_1$ is obtained. Combining the information collected from multiple repetitions of such noninvasive measurements ultimately allows one to obtain the dynamic correlation $C$ with high accuracy. We will present a more detailed summary of the NIMP in Sec.~\ref{sec:recap}.

The original derivation of the NIMP is valid only for correlations of single-site observables, but we show in Sec.~\ref{sec:arbObs} that the noninvasive measurement procedure is in fact valid for arbitrary (multi-site) observables. Dynamic correlations are in general complex, so that two separate implementations of the NIMP are required; One to measure $\rep C$ and another to measure $\imp C$. In Sec.~\ref{sec:simCouple} we discuss a variation of the NIMP in which two noninvasive measurements are performed simultaneously at $t_1$. We show that this allows us to extract both the real and the imaginary part of $C$ from a single set of measurements.

In \cite{Uhrich_etal} (Secs.~VII--VIII) it was shown that dynamic correlations of single-site \mbox{spin-$1/2$} observables can be measured without making use of ancilla degrees of freedom: $\imp C$ can be obtained by performing a local rotation at $t_1$, followed by a projective measurement at $t_2$. $\rep C$ can be obtained by projectively measuring the target both at $t_1$ and at $t_2$, and surprisingly measurement backaction is of no concern in this case. To gain a better understanding of when noninvasive measurements are necessary we use the framework of positive operator-valued measures (POVM) in Sec.~\ref{sec:povm} to analyse the effect that the coupling and measuring of an ancilla has on the target system during the course of the NIMP. This analysis reveals that, in general, noninvasive measurements are necessary to measure dynamic correlations, but that a special class of observables---of which single-site spin-$1/2$ observables are an example---exists for which the effect of the ancilla-based measurement is equivalent to a local rotation (when measuring $\imp C$) or a local projection (when measuring $\rep C$).

\section{Recap of the NIMP}
\label{sec:recap}

A detailed derivation of the noninvasive measurement protocol is given in Sec.~II of \cite{Uhrich_etal}, and we only summarise the pertinent steps here. The NIMP is derived in the context of lattice spin systems consisting of $N$ spin-$s\in\NN/2$ degrees of freedom (no assumptions are made about lattice geometry, size or dimension). The goal is to estimate dynamic correlations $C=\matel{\psi}{S_i^a(t_1)S_j^b(t_2)}{\psi} \in \CC$ of single-site spin-$s$ observables. Here $\ket{\psi}$ is an arbitrary initial state of the lattice at time $t=0$, and the system Hilbert space is $\mathscr{H}_\text{S} = (\CC^{2s+1})^{\otimes N}$. Observable $S_i^a(t_1)=U^\dagger(t_1)S_i^a U(t_1)$ is the $a$-component of a spin-$s$ operator, evaluated at time $t_1\geq 0$ in the Heisenberg picture, with $a\in\{x,y,z\}$ and support $\text{supp}(S_i^a)=\{i\}$. $S_j^b(t_2)$ is defined similarly, only with $t_2>t_1$. The system dynamics $U(t)$ is generated by a Hamiltonian $H_\text{s}$ which may have arbitrary time-dependence.
In what follows we will refer to this spin-$s$ lattice system as the \emph{target}.

As outlined in the introduction, the noninvasive measurement of the target at $t_1$ requires an ancilla system. We choose this ancilla to be a single spin-$s$ degree of freedom with Hilbert space $\mathscr{H}_\text{A}=\CC^{2s+1}$, and assume that its initial state $\ket{\phi}\in\mathscr{H}_\text{A}$ can be prepared independently from that of the target. The combined ancilla--target state at $t=0$ is then a product state
$\ket{\Psi}=\ket{\phi}\otimes\ket{\psi}\equiv\ket{\phi,\psi} \in \mathscr{H}_\text{A} \otimes \mathscr{H}_\text{S}$. Time evolution of $\ket{\Psi}$ from $t=0$ to $t_1\geq 0$ is generated by $H_\text{s}$, which acts nontrivially only on $\mathscr{H}_\text{S}$ so that $\ket{\Psi(t_1)}=[\mathds{1}_\text{A} \otimes U(t_1)]\ket{\phi,\psi} \equiv \ket{\phi,\psi(t_1)}$.

The noninvasive measurement at $t_1$ is achieved by entangling the target with the ancilla, and subsequently measuring the ancilla projectively. The entanglement is generated for a time $\lambda$ by a coupling Hamiltonian $H_\text{c}= B \otimes S^a_i$ which acts on the joint ancilla--target Hilbert space $\mathscr{H}_\text{A} \otimes \mathscr{H}_\text{S}$. The optimal choice for $B$ will be determined in \eqref{e:wmcfFinal}. An essential assumption of the NIMP is that the ancilla--target coupling unitary
\begin{equation}\label{e:couple}
\couple(\lambda) = \exp(-i \lambda H_\text{c}) \simeq \mathds{1} - i\lambda B \otimes S^a_i 
\end{equation}
can be approximated to linear order in $|\lambda| \lVert H_\text{c} \rVert$ (here, and in what follows, we use units where $\hbar=1$). 
We assume, that $H_c$ is bounded, so that $|\lambda| \lVert B \otimes S^a_i \rVert \ll 1$ is achieved when the coupling time $\lambda$ is sufficiently small i.e.\ when $|\lambda| \ll 1$. In \eqref{e:couple} and in what follows we use the symbol $\simeq$ to denote validity up to linear order in $\lambda$.
The ancilla--target entanglement allows one to extract information about the target at $t_1$ by projectively measuring the ancilla at $t_1$, once the coupling is completed. Appendix C of \cite{Uhrich_etal} shows that the ancilla measurement may also be deferred to $t_2$. For the sake of brevity we use this deferred measurement approach here: Once the ancilla--target coupling at $t_1$ is completed, the target is time-evolved to $t_2$ under $H_\text{s}$, yielding
\begin{equation}\label{e:Psit2}
	\ket{\Psi(t_2)} = [\mathds{1}_\text{A} \otimes U(t_2-t_1)] \couple(\lambda) \ket{\phi,\psi(t_1)} \simeq \ket{\phi,\psi(t_2)} - i \lambda B \ket{\phi} \otimes  U(t_2)  S^a_i (t_1) \ket{\psi}.
\end{equation}
At $t_2$, projective measurements of the ancilla and the spin at site $j$ are performed. The relevant measurement bases are chosen to be the eigenbasis of the observables correlated in $C$: The ancilla is measured in the eigenbasis of ${S^a=\sum_{m_a =-s}^s m_a\ket{m_a}\!\bra{m_a}}$ (which appears in $C$ at $t_1$), and lattice site $j$ is measured in the eigenbasis of $S_j^b=\sum_{m_b =-s}^s m_b \ket{m_b}\!\bra{m_b}$ (which appears in $C$ at $t_2$). The measurement outcome is any of the $(2s+1)^2$ combinations of eigenvalues $m_a,m_b\in\mathscr{S}=\{-s,-s+1,\ldots,s\}$. The probability to measure eigenvalues $(m_a,m_b)\in \mathscr{S} \times \mathscr{S}$ is given by Born's rule as
\begin{equation}\label{e:jointProb}
	\begin{split}
		P_{m_a,m_b} =& \matel{\Psi(t_2)}{\left( \ket{m_a}\!\bra{m_a} \otimes \ket{m_b}\!\bra{m_b} \right)}{\Psi(t_2)} \\
		\simeq& \innprod{\phi}{m_a}\!\innprod{m_a}{\phi}\matel{\psi}{U^\dagger(t_2)\ket{m_b}\!\bra{m_b}U(t_2)}{\psi} \\
		&-2\lambda \impart{\matel{\phi}{B}{m_a}\!\innprod{m_a}{\phi}\matel{\psi}{ S^a_i (t_1) U^\dagger(t_2) \ket{m_b}\!\bra{m_b} U(t_2)}{\psi} },
	\end{split}
\end{equation}
where the projector $\ket{m_a}\bra{m_a}$ acts on $\mathscr{H}_\text{A}$ and the  projector $\ket{m_b}\bra{m_b}$ acts only on the $j$th spin of the target.
Using \eqref{e:jointProb} we correlate the measured eigenvalues as
\begin{equation}\label{e:wmcf}
	\begin{split}
	\wmcf(t_1,t_2) =& \sum_{ m_a, m_b \in \mathscr{S} } m_a m_b P_{m_a , m_b} \\
	\simeq& \expval{S^a}_{\phi} \expval{S_j^b(t_2)}_{\psi} - 2 \lambda \imp \left( \expval{B S^a}_{\phi} \matel{\psi}{  S^a_i(t_1) S_j^b(t_2) }{\psi} \right),
	\end{split}	
\end{equation}
where we have absorbed the summations via the spectral representations of $S^a$ and $S_j^b$.
The last term in \eqref{e:wmcf} contains the desired correlation $C$. The first term is thus an unwanted constant and can be eliminated by choosing $\ket{\phi}$ such that $\expval{S^a}_\phi=0$. This is achieved with any normalised state 
\begin{equation}\label{e:phi}
	\ket{\phi}=\sum_{m_a\in \mathscr{S}}r_{m_a} e^{i\theta_{m_a}} \ket{m_a} 
\end{equation}
that satisfies $\theta_{m_a} \in \RR$ and $r_{m_a}=r_{-m_a}$ for all $m_a \in \mathscr{S}$. For simplicity we choose $r_{m_a} e^{i\theta_{m_a}}=(2s+1)^{-1/2}$, in which case
\begin{equation}\label{e:wmcfFinal}
	\wmcf(t_1,t_2) = 
	\begin{cases}
		\wmcf^{(1)} \simeq -2\lambda f^{(1)} \imp{\matel{\psi}{ S^a_i(t_1) S_j^b(t_2) }{\psi}} & \text{for $B=B^{(1)}=S^a$}, \\
		\wmcf^{(2)} \simeq -2\lambda f^{(2)} \rep{\matel{\psi}{ S^a_i(t_1) S_j^b(t_2) }{\psi}} & \text{for $B=B^{(2)}=\frac{i}{2}\left(S_a^--S_a^+\right)$}.
	\end{cases}
\end{equation}
Here $S_a^{\pm}$ denote the spin-$s$ ladder operators with respect to the eigenbasis of $S^a$.
The prefactors $f^{(n)}$ stem from the $\expval{B^{(n)}S^a}_\phi$ term in \eqref{e:wmcf}. For our choice of $\ket{\phi}$ [see below \eqref{e:phi}] we have
\begin{equation}\label{e:deff}
	\begin{split}
		f^{(1)}=&\frac{1}{2s+1}\sum_{m_a\in\mathscr{S}}m_a^2=\frac{2s^3+3s^2+s}{3(2s+1)}, \\
		f^{(2)}=&\frac{i}{2s+1}\sum_{m_a,m'_a\in\mathscr{S}} m_a\matel{m_a}{B^{(2)}}{m'_a}=\frac{1}{2(2s+1)}\sum_{m_a=-s}^{s-1} c_+(s,m_a),
	\end{split}
\end{equation}
where $c_+(s,m_a)=\sqrt{s(s+1)-m_a(m_a+1)}$.

Equation \eqref{e:wmcfFinal} is the main result of the NIMP. It proves that probing the target at $t_1$ with a noninvasive measurement (i.e., in the limit $|\lambda|\ll 1$) allows one to obtain both the real and the imaginary parts of $C$. This requires two separate implementations of the NIMP---one for either choice of $B$ in \eqref{e:wmcfFinal}---from which one can then construct the complex-valued dynamic correlation as
\begin{equation}\label{e:reconstruct}
	C^\lambda \simeq -\frac{1}{2\lambda} \left(\frac{\wmcf^{(2)}(t_1,t_2)}{f^{(2)}} +i \frac{\wmcf^{(1)}(t_1,t_2)}{f^{(1)}} \right).
\end{equation}
To do so one must estimate the probabilities $P_{m_a,m_b}$ \eqref{e:jointProb} by repeatedly preparing the ancilla--target state $\ket{\phi,\psi}$ and executing the above protocol $n$ times. In this way a sample of $n$ measured eigenvalue pairs is obtained, and in principle the relative frequency $n_{m_a m_b}/n$ with which a given eigenvalue pair is measured converges to $P_{m_a,m_b}$ as $n\to\infty$ and in the limit of infinitesimal $\lambda$. In the realistic situation of finite $n$ and $\lambda$, the relative frequencies will deviate from $P_{m_a,m_b}$. These deviations subsequently propagate into correlations $\wmcf^{(1)},\wmcf^{(2)}$, so that $C^\lambda$ will deviate from the true value of $C$.
A detailed error analysis is presented in Sec.~III of \cite{Uhrich_etal}, where it is shown that for a given number of measurements $n$ there exists an optimal coupling time $\lambda^*$ for which the overall deviation of $C^\lambda$ from $C$ is minimised.

\section{Generalisations}
\label{sec:gen}

\subsection{Arbitrary observables and ancilla spin number}
\label{sec:arbObs}
The above results were derived in Ref.~\cite{Uhrich_etal} for dynamic correlations between single-site spin operators $S_i^a$ and $S_j^b$.
We now generalise the NIMP to dynamic correlations $C=\matel{\psi}{O_1(t_1) O_2(t_2)}{\psi}$ of arbitrary (multi-site) spin-$s$ observables $O_1$ and $O_2$. A relevant example is the lattice magnetisation $\frac{1}{N}\sum_{n=1}^N S_n^a$, or sublattice magnetisations. The target system, as introduced in Sec.~\ref{sec:recap}, is an arbitrary lattice of spin-$s$ degrees of freedom.
As before, we use a single spin degree of freedom as the ancilla. In contrast to Sec.~\ref{sec:recap} we do not assume the ancilla's spin quantum number $\zeta$ to match that of the target's individual spins i.e.\ $\zeta=s$ is not required. Analogous to Sec.~\ref{sec:recap}, the coupling Hamiltonian---which generates the weak ancilla--target coupling $\couple(\lambda)$ at $t_1$---is chosen as $H_{\text{c}}=B\otimes O_1$. The entangled ancilla--target state at $t_2$ is thus formally the same as \eqref{e:Psit2}, only with $S^a_i$ replaced by $O_1$. 
Using the deferred measurement approach, we projectively measure at time $t_2$ a suitably chosen observable on the ancilla Hilbert space $\mathscr{H}_A$ and another one acting nontrivially only on the support of $O_2$. A natural choice for the latter measurement is the eigenbasis of ${O_2=\sum_{o}e_o\Pi^o}$, where $\Pi^o$ denotes the projector onto the eigenspace corresponding to the eigenvalue $e_o$ of $O_2$, which may be degenerate. For the ancilla measurement we choose the measurement basis to be the eigenbasis of the \mbox{spin-$\zeta$} observable $S^\alpha=\sum_{m_\alpha=-\zeta}^\zeta m_\alpha \ket{m_\alpha}\!\bra{m_\alpha}$, where $\alpha\in \{x,y,z\}$. The probability for measuring eigenvalues $(m_\alpha,e_o)$ is then
\begin{multline}
		P_{m_\alpha, e_o} = \matel{\Psi(t_2)}{(\ket{m_\alpha}\!\bra{m_\alpha}\otimes \Pi^o)}{\Psi(t_2)} \\
		\simeq \innprod{\phi}{m_\alpha}\innprod{m_\alpha}{\phi}\matel{\psi}{\Pi^o(t_2)}{\psi} -2\lambda \impart{\matel{\phi}{B}{m_\alpha}\innprod{m_\alpha}{\phi}\matel{\psi}{O_1(t_1)\Pi^o(t_2)}{\psi} },
	\end{multline}
where $\Pi^o(t_2)=U^\dagger(t_2) \Pi^o U(t_2)$.
Correlating the measured eigenvalues as in \eqref{e:wmcf} we obtain
\begin{equation}\label{e:32}
\begin{split}
\wmcf(t_1,t_2) &=\sum_{m_\alpha,o} m_\alpha e_o P_{m_\alpha, e_o}\\ 
&\simeq \expval{S^\alpha}_{\phi} \expval{O_2(t_2)}_{\psi} - 2 \lambda \imp \left(\expval{ B S^\alpha}_{\phi}\matel{\psi}{ O_1(t_1) O_2(t_2) }{\psi}\right).
\end{split}
\end{equation}
The first term in the second line of \eqref{e:32} can be eliminated by choosing $\ket{\phi}$ to satisfy \eqref{e:phi} so that ${\expval{S^\alpha}_{\phi}=0}$. As in Sec.~\ref{sec:recap} [see below \eqref{e:phi}] we choose
\begin{equation}
\ket{\phi}=\sum_{m_\alpha \in \mathscr{S}} \ket{m_\alpha}/\sqrt{2\zeta+1}
\end{equation}
as an equal superposition of the ancilla measurement basis states $\ket{m_\alpha}$. The choices of $B$ which yield $\rep C$ and $\imp C$, respectively, are analogous to those in Eq.~\eqref{e:wmcfFinal} for the single-site NIMP,
\begin{equation}\label{e:genNIMPc}
\wmcf(t_1,t_2) \simeq 
\begin{cases}
-2\lambda f^{(1)} \imp{\matel{\psi}{ O_1(t_1) O_2(t_2) }{\psi}} & \text{for $B=B^{(1)}=S^\alpha$}, \\
-2\lambda f^{(2)} \rep{\matel{\psi}{ O_1(t_1) O_2(t_2) }{\psi}} & \text{for $B=B^{(2)}=\frac{i}{2}\left(S_\alpha^--S_\alpha^+\right)$},
\end{cases}
\end{equation}
where $f^{(n)}$ is defined as in \eqref{e:deff}. This result is formally the same as \eqref{e:wmcfFinal}, but is a generalisation thereof: 
It is valid for correlations of arbitrary multi-site observables $O_1$ and $O_2$, and provides some freedom of choice for the measurement basis of the ancilla measurement, as well as the size of the ancilla spin $\zeta$.

In Sec.~\ref{sec:recap} we made the physically intuitive assumption that the measurement basis for the ancilla measurement should be the eigenbasis of the observable $S_i^a$ which is to be correlated at $t_1$. This assumption requires $\text{dim}(\mathscr{H}_\text{A})=\CC^{2s+1}$ and thus restricts the ancilla's spin quantum number to match that of the target's single-site degrees of freedom, i.e.\ ${\zeta=s}$. Equation \eqref{e:genNIMPc}, however, holds for any $\zeta \in \NN/2$ and ${\alpha \in \{x,y,z\}}$. For any given target system and arbitrary observables $O_1$ and $O_2$, the theoretically simplest choice of ancilla spin and measurement basis is $\zeta=1/2$ and $\alpha=z$, in which case
\begin{equation}\label{e:genEG}
2\wmcf(t_1,t_2) \simeq 
\begin{cases}
-\lambda \imp{\matel{\psi}{ O_1(t_1) O_2(t_2) }{\psi}} & \text{for $B=\sigma^z/2$}, \\
-\lambda \rep{\matel{\psi}{ O_1(t_1) O_2(t_2) }{\psi}} & \text{for $B=\sigma^y/2$},
\end{cases}
\end{equation}
where $\sigma^\alpha$ denotes the $\alpha$-component of the Pauli vector operator.
From an experimental viewpoint the ideal choice for ancilla spin number and measurement basis will depend on the capabilities of the set-up. Implementing the NIMP in linear ion-traps (discussed in \cite{Uhrich_etal}, Sec.~VI) can be achieved by designating one of the trapped ions as the ancilla. Typically a single ionic species is trapped and a single transition is selected as pseudo-spin degree of freedom, so that naturally $\zeta=s$ (for reviews on ion trapping see for instance \cite{Steane1997,Ozeri2011,James97}). More general set-ups, including mixed ion species and/or addressing of several transitions, are however also possible or conceivable \cite{Home2013,Wright2016}, and in this case ancillas with spin $\zeta\neq s$ may be implemented.

\subsection{Simultaneous noninvasive measurements}
\label{sec:simCouple}
Measuring a complex-valued dynamic correlation $C$ requires two sets of measurements: one for $\rep C$ and one for $\imp C$. This is evident in the NIMP where one requires two distinct noninvasive measurements of the target: $\rep C$ is obtained with ancilla--target coupling Hamiltonian $H^{(2)}_\text{c}$, and $\imp C$ is obtained with a different coupling Hamiltonian $H^{(1)}_\text{c}$. To extract $C$ as in \eqref{e:reconstruct}, our approach has thus far been to implement only one of these noninvasive measurements at a time, so that two separate implementations of the NIMP were required to measure $C=\matel{\psi}{ O_1(t_1) O_2(t_2) }{\psi}$.

Here we discuss an alternative approach which requires only a single measurement procedure. The key idea is to perform the two noninvasive measurements of the separate NIMP implementations simultaneously, in the sense that two ancilla degrees of freedom are coupled to the target at the early time $t_1$. We denote the two ancilla initial states as $\ket{\phi_1}$ and $\ket{\phi_2}$ and assume that they form a product state among each other and with the target initial state $\ket{\psi}$ at $t=0$,
\begin{equation}
	\ket{\Psi} = \ket{\phi_1}\otimes \ket{\phi_2}\otimes \ket{\psi} \equiv \ket{\phi_1,\phi_2,\psi} \in \mathscr{H}_{\text{A}_1}\otimes\mathscr{H}_{\text{A}_2}\otimes\mathscr{H}_\text{S}.
\end{equation}
For simplicity we choose both ancillas to be spin-$1/2$ degrees of freedom, but, as shown in Sec.~\ref{sec:arbObs}, this restriction can be lifted. At $t_1$, ancillas $1$ and $2$ are coupled to the target via coupling unitaries
\begin{equation}\label{e:simCouple}
	\begin{split}
		\couple_1(\lambda_1)=&e^{-i\lambda_1 H_{\text{c}1}} \quad\text{with $H_{\text{c}1} = B_1\otimes\mathds{1}_2\otimes O_1$},\\
		\couple_2(\lambda_2)=&e^{-i\lambda_2 H_{\text{c}2}} \quad\text{with $H_{\text{c}2} = \mathds{1}_1\otimes B_2\otimes O_1$},
	\end{split}
\end{equation}
respectively. As before, we assume that the coupling times satisfy $|\lambda_1|,|\lambda_2|\ll 1$ so that (for $k=1,2$) ${\couple_k(\lambda_k) \simeq \mathds{1}-i\lambda_k  H_{\text{c}k}}$.
Note that $\left[ H_{\text{c}1}, H_{\text{c}2}\right]=0$, and therefore the order in which the ancillas are coupled to the target is irrelevant. Using the deferred measurement approach, both ancillas and the target are projectively measured at $t_2$. The combined ancilla--target state at $t_2$ is
\begin{equation}
	\begin{split}
		\ket{\Psi(t_2)} =& [\mathds{1}_1\otimes\mathds{1}_2\otimes U(t_2-t_1)]\couple_1(\lambda_1)\couple_2(\lambda_2) \ket{\phi_1,\phi_2,\psi(t_1)} \\
		\simeq& \ket{\phi_1,\phi_2,\psi(t_2)} -i\lambda_1 B_1\ket{\phi_1}\ket{\phi_2}U(t_2)O_1(t_1)\ket{\psi} \\
		&-i\lambda_2 \ket{\phi_1}B_2\ket{\phi_2}U(t_2)O_1(t_1)\ket{\psi} -\lambda_1\lambda_2 B_1\ket{\phi_1}B_2\ket{\phi_2}U(t_2)(O_1 )^2(t_1)\ket{\psi} .
	\end{split}
\end{equation}
The basis for the target measurement is the eigenbasis of $O_2$ (as in Sec.~\ref{sec:arbObs}). For simplicity (see also Sec.~\ref{sec:arbObs}) we choose the measurement basis for both spin-$1/2$ ancillas to be the eigenbasis of $S^z=\sigma^z/2$, and we denote the measured eigenvalues for the two ancillas as $m_1,m_2\in\{-1/2,+1/2\}$. The probability of measuring eigenvalues $(m_1,m_2,e_o)$ is then
\begin{equation}\label{e:simProbJoint}
		P_{m_1,m_2,e_o} = \matel{\Psi(t_2)}{(\ket{m_1}\!\bra{m_1}\otimes\ket{m_2}\!\bra{m_2}\otimes\Pi^o)}{\Psi(t_2)},
	\end{equation}
from which we obtain two marginal distributions
\begin{equation}\label{e:marg}
	\begin{split}
			P_{m_1,e_o} =& \sum_{m_2=\pm 1/2} P_{m_1,m_2,e_o} = \matel{\Psi(t_2)}{(\ket{m_1}\!\bra{m_1}\otimes\mathds{1}_2\otimes\Pi^o)}{\Psi(t_2)}, \\
			P_{m_2,e_o} =& \sum_{m_1=\pm1/2} P_{m_1,m_2,e_o} = \matel{\Psi(t_2)}{(\mathds{1}_1\otimes\ket{m_2}\!\bra{m_2}\otimes\Pi^o)}{\Psi(t_2)}.
	\end{split}
\end{equation}
We can extract both $\rep C$ and $\imp C$ from these marginal distributions as follows.

\paragraph{Imaginary part.}
We use $P_{m_1,e_o}$ to correlate the eigenvalues obtained from measuring ancilla $1$ and the target as
\begin{equation}\label{e:wmcfIm}
	\begin{split}
		\wmcf_1 =& \sum_{m_1,o}m_1 e_o P_{m_1,e_o} = \matel{\Psi(t_2)}{(S^z\otimes\mathds{1}_2\otimes O_2)}{\Psi(t_2)} \\
		\simeq& \expval{S^z}_{\phi_1}\expval{O_2(t_2)}_{\psi} -2\lambda_1\imp\!\left(\expval{B_1S^z}_{\phi_1} \matel{\psi}{O_1(t_1)O_2(t_2)}{\psi}\right)\\
		& -2\lambda_2 \expval{S^z}_{\phi_1} \expval{B_2}_{\phi_2} \imp\left(\matel{\psi}{O_1(t_1)O_2(t_2)}{\psi}\right)\\
		&+2\lambda_1\lambda_2 \expval{B_2}_{\phi_2} \rep\!\left(\expval{B_1S^z}_{\phi_1}\left( \expval{O_1(t_1)O_2(t_2)O_1(t_1)}_\psi - \expval{(O_1)^2(t_1) O_2(t_2)}_\psi \right) \right).
	\end{split}
\end{equation}
The third term is proportional to the desired quantity $\imp C$, but the second term is more useful since $\imp C$ can easily be isolated: If $\ket{\phi_1}$ satisfies \eqref{e:phi}, $\expval{S^z}_{\phi_1}=0$ and the first and third term of \eqref{e:wmcfIm} vanish. For simplicity we choose $\ket{\phi_1}=(\ket{+1/2}+\ket{-1/2})/\sqrt{2}$ [note that other choices will only lead to different prefactors in \eqref{e:simIm}].
The last term of \eqref{e:wmcfIm} can also be eliminated by choosing $\ket{\phi_2}$ such that $\expval{B_2}_{\phi_2} =0$. Then \eqref{e:wmcfIm} reduces to
\begin{equation}\label{e:simIm}
		\wmcf_1 \simeq - \frac{\lambda_1}{2}\imp C \quad\text{for $\expval{S^z}_{\phi_1} =0$, $\expval{B_2}_{\phi_2} =0$, and $B_1=S^z$.}
\end{equation}

\paragraph{Real part.} 
Analogous to \eqref{e:wmcfIm}, we use $P_{m_2,e_o}$ to correlate the eigenvalues obtained from measuring ancilla $2$ and the target. Keeping in mind that we have already chosen $B_1=S^z$ and $\ket{\phi_1}=(\ket{+1/2}+\ket{-1/2})/\sqrt{2}$, we obtain
\begin{equation}\label{e:wmcfRe}
\begin{split}
\wmcf_2 =& \sum_{m_2,o}m_2 e_o P_{m_2,e_o} = \matel{\Psi(t_2)}{(\mathds{1}_1\otimes S^z\otimes O_2)}{\Psi(t_2)} \\
\simeq& \expval{S^z}_{\phi_2}\expval{O_2(t_2)}_{\psi} -2\lambda_2\impart{\expval{B_2S^z}_{\phi_2} \matel{\psi}{O_1(t_1)O_2(t_2)}{\psi}} 
.
\end{split}
\end{equation}
To extract $\rep C$ from \eqref{e:wmcfRe} we choose $\ket{\phi_2}=\ket{\phi_1}$ so that $\ket{\phi_2}$ also satisfies \eqref{e:phi} and thus $\expval{S^z}_{\phi_2}=0$. This eliminates the first term in \eqref{e:wmcfRe}. What remains is to choose $B_2$ such that $\expval{B_2S^z}_{\phi_2}$ is purely imaginary and such that, simultaneously, the requirement $\expval{B_2}_{\phi_2} =0$ from \eqref{e:simIm} is satisfied. Having chosen $\ket{\phi_2}=\ket{\phi_1}=(\ket{+1/2}+\ket{-1/2})/\sqrt{2}$, the choice ${B_2=S^y=\sigma^y/2}$ meets both conditions, leading to
\begin{equation}\label{e:simRe}
	\wmcf_2 \simeq - \frac{\lambda_2}{2}\rep C .
\end{equation}
In this derivation we have followed a similar strategy as in Secs.~\ref{sec:recap} and \ref{sec:arbObs}, only with the new technical ingredient of using two marginal distributions \eqref{e:simProbJoint}--\eqref{e:marg} to construct correlations $\wmcf_1$ and $\wmcf_2$. The choice of operators $B_1=S^z$ and $B_2=S^y$, which allows us to extract $\imp C$ \eqref{e:simIm} and $\rep C$ \eqref{e:simRe}, respectively, are the same as for the generalised NIMP of Sec.~\ref{sec:arbObs}. 
The initial ancilla states must again satisfy \eqref{e:phi} (so that ${\expval{S^z}_{\phi}=0}$), and we have an additional constraint $\expval{B_2}_{\phi_2}=0$ on $\ket{\phi_2}$. With $B_2=S^y$, this too is satisfied by our choice $\ket{\phi_{1}}=\ket{\phi_2}=(\ket{+1/2}+\ket{-1/2})/\sqrt{2}$. In App.~\ref{app:proofB2} we derive the general class of ancilla initial states that may be used for the simultaneous noninvasive measurements. Physically relevant examples in that class include equal superpositions of basis states as well as spin-coherent states.

\section{Noninvasive measurements within the framework of positive operator-valued measures}
\label{sec:povm}

The main ingredient of the NIMP, and also of its generalisations in Sec.~\ref{sec:gen}, is the coupling of an ancilla to the target at time $t_1$, and the subsequent measurement of that ancilla. 
During the coupling procedure, ancilla and target become entangled so that the subsequent measurement of the ancilla also affects the target state.
To understand this effect on the target we now employ the formalism of \emph{positive operator-valued measures} (POVMs). 
For any ancilla-based measurement, it is always possible to described its effect on the target system by means of a POVM  (conversely, it is also always possible to express a POVM in terms of an ancilla-based measurement; see Sec.~$9.6$ of \cite{AsherPeres}). A POVM is conveniently described through a set of so-called \emph{Kraus operators}\/ or \emph{measurement operators}\/ $M_l$, which act on the target Hilbert space $\mathscr{H}_\text{S}$ only. The index $l$ labels the measurement basis $\{\ket{l}\}$ of the ancilla, and the Kraus operators satisfy $\sum_l M_lM^\dagger_l=\mathds{1}$ in order to guarantee a probability interpretation of the measurement outcomes. Introductions on Kraus operators and POVM can be found in \cite{Jacobs,NielsenChuang,AsherPeres}.

\subsection{Correlations of general observables.}
The goal of this section is to derive the POVM, in the sense of a mapping acting on the target state only, that describes the effect of the generalised NIMP of Sec.~\ref{sec:arbObs}. We want the POVM to capture only the effect due to the ancilla--target coupling and the measurement on the ancilla, but not to include the target time-evolution from $t_1$ to $t_2$. For this reason we switch to a protocol where the ancilla is measured at $t_1$ right after having been coupled to the system (it has been shown in Appendix~C of \cite{Uhrich_etal} that this approach and the deferred measurement approach used in Sec.~\ref{sec:gen} of the present paper lead to identical results for the dynamic correlation functions). For now we do not specify the spin quantum number $\zeta\in\NN/2$ of the ancilla spin. At $t_1$ the ancilla is measured in the eigenbasis of $S^\alpha$, and the (un-normalised) post-measurement ancilla--target state corresponding to a measured eigenvalue $m_\alpha \in\{-\zeta,\ldots,\zeta\}$ is
\begin{equation}
	\ket{\Psi_{m_\alpha}(t_1)} = (\ket{m_\alpha}\!\bra{m_\alpha}\otimes\mathds{1}_\text{S}) \couple(\lambda)\ket{\phi,\psi(t_1)}.
\end{equation}
Tracing over the ancilla degrees of freedom yields the reduced density matrix of the target,
\begin{equation}\label{e:rho}
	\begin{split}
		\rho_{m_\alpha}(t_1)&= \Tr_{ \text{A}}\left[\ket{m_\alpha}\!\bra{m_\alpha}\mathscr{U}(\lambda)\ket{\phi,\psi(t_1)}\!\bra{\phi,\psi(t_1)} \mathscr{U}^\dagger(\lambda)  \ket{m_\alpha}\!\bra{m_\alpha} \right] \\
		&=M_{m_\alpha} \rho(t_1)M^\dagger_{m_\alpha},  
	\end{split}
\end{equation}
where $\Tr_{ \text{A}}$ denotes the partial trace over the ancilla Hilbert space $\mathscr{H}_\text{A}$, and $\rho(t_1)=\ket{\psi(t_1)}\!\bra{\psi(t_1)}$ is the density operator of the target after time-evolution up to $t_1$, but before the ancilla coupling is switched on. The second line of Eq.~\eqref{e:rho} is essentally a definition of the Kraus operators in the context of the NIMP, requiring that the coupling and subsequent measurement of the ancilla maps the target state $\rho(t_1)$ onto $M_{m_\alpha} \rho(t_1)M^\dagger_{m_\alpha}$ if the outcome of the ancilla measurement was $m_\alpha$. The Kraus operators $M_{m_\alpha}$ are given by
\begin{equation}\label{e:kraus}
	\begin{split}
		M_{m_\alpha} &= \matel{m_\alpha}{\mathscr{U}(\lambda)}{\phi}
		= \sum_{n=0}^\infty \frac{(-i\lambda)^n}{n!} \matel{m_\alpha}{B^n}{\phi}O_1^n \\
		&= \sum_{n=0}^\infty \frac{(-i\lambda)^n}{n!}\sum_{m'_\alpha = -\zeta}^\zeta c_{m'_\alpha} \matel{m_\alpha}{B^n}{m'_a} O_1^n,
	\end{split}
\end{equation}
where we have used $\ket{\phi}=\sum_{m'_\alpha=-\zeta}^\zeta c_{m'_\alpha} \ket{m'_\alpha}$ to get to the second line.
Equation \eqref{e:kraus} gives the general form of the Kraus operators describing the effect of the ancilla-based NIMP on the target system only. For general ancilla spin quantum number $\zeta$ the Kraus operators are not easy to interpret but, as we will see, they become reasonably suggestive in the special case $\zeta=1/2$, which we will specialise to from here onwards. Measuring in the eigenbasis $\{\ket{\pm}\}$ of $S^z=\frac{1}{2}\sigma^z$, we denote the corresponding Kraus operators as $M_\pm$. The ancilla initial state, subject to the constraint \eqref{e:phi}, is parametrised as 
\begin{equation}\label{e:phiKraus}
	\ket{\phi}=(e^{i\theta_+}\ket{+}+e^{i\theta_-}\ket{-})/\sqrt{2} \quad\text{with $\theta_\pm \in \RR$}.
\end{equation}  

\paragraph{Imaginary part.}
In Sec.~\ref{sec:gen} we found that we need to choose $B=S^z$ in the target--ancilla coupling Hamiltonian $H_\text{c}$ in order to obtain the imaginary part $\imp C$ of the desired dynamic correlation function. Plugging this choice into \eqref{e:kraus} yields the Kraus operators
\begin{equation}\label{e:krausIm}
		M^\text{I}_{\pm} = \sum_{n=0}^\infty \frac{(-i\lambda)^n}{n!} \matel{\pm}{(S^z)^n}{\phi}O_1^n  
		= \innprod{\pm}{\phi} e^{\mp i\lambda O_1 /2} ,
\end{equation}
where we have used $\bra{\pm}(S^z)^n=(\pm1/2)^n$. The superscript $\text{I}$ in \eqref{e:krausIm} refers to the {\em imaginary} part of $C$.
Up to the prefactor, \eqref{e:krausIm} is a unitary time-evolution of the target, generated by $O_1$ during a time $\pm\lambda /2$ (or, equivalently, by $\pm O_1$ during a time $\lambda /2$). The unitary form of the Kraus operators $M_\pm^\text{I}$ for $\imp C$ is a consequence of the ancilla measurement basis used in the NIMP being also the eigenbasis of the operator $B=S^z$ in the ancilla--target coupling. Unitarity suggests that we can measure $\imp C$ without invoking an ancilla, but rather by directly probing the unitarily evolved target state
\begin{equation}
	\ket{\psi(t_2,\pm\lambda)} = U(t_2-t_1)e^{-i(\pm\lambda) O_1}U(t_1)\ket{\psi}
\end{equation}
in the eigenbasis of $O_2$ at $t_2$. Indeed, to first order in $\lambda$ we have
\begin{equation}
	\matel{\psi(t_2,\pm\lambda)}{O_2}{\psi(t_2,\pm\lambda)} \simeq \matel{\psi}{O_2(t_2)}{\psi} \mp2\lambda\impart{\matel{\psi}{O_1(t_1)O_2(t_2)}{\psi}},
\end{equation}
from which we can extract $\imp C$ by taking the difference
\begin{equation}\label{e:imRot}
	\matel{\psi(t_2,-\lambda)}{O_2}{\psi(t_2,-\lambda)} -	\matel{\psi(t_2,+\lambda)}{O_2}{\psi(t_2,+\lambda)} \simeq 4\lambda \impart{\matel{\psi}{O_1(t_1)O_2(t_2)}{\psi}}
\end{equation}
between expectation values with forward and backward evolved states.
We emphasise that no ancilla is required, as all operations are performed directly on the target. A physically relevant scenario is when $O_1$ is the total magnetisation ${O_1=\sum_{n=1}^N S_n^a/N}$ of the target. In this case the Kraus operators \eqref{e:krausIm} are global rotations $\prod_{n=1}^N e^{\mp i \theta S_n^a/N}$ (with $\lambda/2=\theta$), and $\imp C$ can be extracted as in \eqref{e:imRot} by requiring $|\theta|\ll1$. For certain experimental implementations, a (local or global) rotation may be significantly easier to implement than coupling an ancilla to the target.

\paragraph{Real part.}
In the previous paragraph we have found conditions under which the Kraus operators $M_\pm^\text{I}$ are unitary, and this has allowed us to find a measurement protocol for $\imp C$ that does not require an ancilla. Similarly, it would be desirable to find conditions under which the Kraus operators $M_\pm^\text{R}$, which describe the noninvasive measurement of $\rep C$ are unitary, as this would then point towards an ancilla-free protocol also for the measurement of $\rep C$. Sec.~\ref{sec:gen} showed that, when using the eigenbasis of $S^z=\sigma^z/2$ as the ancilla measurement basis, $\rep C$ is obtained by choosing $B=S^y=\sigma^y/2$ in the coupling operator of the NIMP. The ancilla measurement basis is thus not the eigenbasis of $B$ in this case, and operators $M_\pm^\text{R}$ do not have an immediate unitary form [in contrast to $M_\pm^\text{I}$ \eqref{e:krausIm}]. However, by choosing $\ket{\phi}$ as an eigenstate of $B$, i.e.\ $B\ket{\phi}=\phi_\text{B}\ket{\phi}$ (with $\phi_\text{B}\in\RR$ since $B^\dagger=B$), the Kraus operators \eqref{e:kraus} become unitary
\begin{equation}
		M_{\pm} = \innprod{\pm}{\phi}e^{-i(\lambda \phi_\text{B})O_1}.
\end{equation}
Equation~\eqref{e:32} however shows that for such an initial ancilla state the NIMP can never yield $\rep C$ since then
\begin{equation}
\begin{split}
\wmcf(t_1,t_2) \simeq \expval{S^z}_{\phi} \left( \expval{O_2(t_2)}_{\psi} - 2 \lambda \phi_\text{B} \imp \matel{\psi}{ O_1(t_1) O_2(t_2) }{\psi}\right).
\end{split}
\end{equation}%
To measure $\rep C$, we must thus ensure that $\ket{\phi}$ is not an eigenstate of $S^y$. The $\ket{\pm}$ states in \eqref{e:phiKraus} may therefore not have a relative phase of $i$, and we choose (without loss of generality) $\ket{\phi}=(\ket{+}+\ket{-})/\sqrt{2}$. We denote the spectral resolution of $O_1$ as $O_1=\sum_\omega e_\omega \Pi^\omega$, where operators $\Pi^\omega$ project onto the (possibly degenerate) eigenspaces of $O_1$. Substituting $B=S^y$ and our choice of $\ket{\phi}$ into \eqref{e:kraus}, one can then show (using $(S^y)^2=\mathds{1}/4$) that the Kraus operators for the noninvasive measurement of $\rep C$ are 
\begin{equation}\label{e:krausRe}
\begin{split}
	M_\pm^\text{R}=&\frac{1}{\sqrt{2}} \left[\cos\left(\frac{\lambda}{2}O_1 \right) \mp \sin\left(\frac{\lambda}{2}O_1\right)\right] = \frac{1}{\sqrt{2}} \sum_\omega \left[ \cos \left( \frac{\lambda e_\omega}{2} \right) \mp \sin\left(\frac{\lambda e_\omega}{2}\right)\right] \Pi^\omega \\
	\simeq& \frac{1}{\sqrt{2}} \sum_\omega \left[ 1 \mp \frac{\lambda e_\omega}{2}\right] \Pi^\omega.
\end{split}
\end{equation}
In contrast to $M_\pm^\text{I}$, the operators $M_\pm^\text{R}$ are not unitary. In fact, \eqref{e:krausRe} shows that when the NIMP is set up so as to measure $\rep C$ (with $|\lambda|\ll1$), the effect on the target state $\ket{\psi(t_1)}$ is a simultaneous projection  onto \emph{all} eigenspaces of $O_1$. It is not clear to us how to achieve the effect of the operators $M_\pm^\text{R}$, for an arbitrary spin-$s$ observable $O_1$, via some combination of unitary transformations and/or projections acting only on the target Hilbert space $\mathscr{H}_\text{S}$. Our present understanding is thus that one must, in general, make use of a noninvasive ancilla-based measurement at $t_1$ in order to measure $\rep \matel{\psi}{O_1(t_1)O_2(t_2)}{\psi}$.

\subsection{Spin-$1/2$ target systems.}
Assuming the target to be a lattice consisting of spin-$1/2$ degrees of freedom, there are certain observables $O_1$ for which Kraus operators $M_\pm^\text{I}$ \eqref{e:krausIm} and $M_\pm^\text{R}$ \eqref{e:krausRe} reduce, respectively, to rotations and projections onto a single eigenspace of $O_1$.

When $O_1$ acts only on a single lattice site $j$, i.e.\ $O_1=\frac{1}{2}\sigma_j^a$, we have $(O_1)^2\propto \mathds{1}_j$, which implies that the Kraus operators $M_\pm^\text{I}$ are linear in $O_1$ for all $\theta$. This negates the need for a linear expansion in $\lambda$ (or $\theta$) as was used in the derivation of \eqref{e:imRot}. Consequently, $\imp C$ can be measured by performing a local rotation (not restricted to small rotation angles) of the target at $t_1$, followed by a projective measurement in the eigenbasis of $O_2$ at $t_2$.

The Kraus operators for $\rep C$ simplify when the spectrum of $O_1$ satisfies
\begin{equation}\label{e:spectrum}
	O_1 =e ( \Pi^1 - \Pi^2) \text{ and } e\in\RR .
\end{equation}
Observables which satisfy this condition are again single-site spin-$1/2$ observables, as well as multi-site tensor products of these.
In this case $M_\pm^\text{R}$ \eqref{e:krausRe} simplifies to
\begin{equation}\label{e:krausReSpecial}
M_\pm^\text{R}= \frac{1}{\sqrt{2}} \left[ \cos \left( \frac{\lambda e}{2} \right) \mp \sin\left(\frac{\lambda e}{2}\right)\right] \Pi^1 + \frac{1}{\sqrt{2}} \left[ \cos \left( \frac{\lambda e}{2} \right) \pm \sin\left(\frac{\lambda e}{2}\right)\right] \Pi^2.
\end{equation}
Equation~\eqref{e:krausReSpecial} is valid for arbitrary $\lambda \in \RR$. Choosing $\lambda=\pi/2e$, \eqref{e:krausReSpecial} reduces to projections onto only a single eigenspace of $O_1$,
\begin{equation}
	M_+^\text{R}=\Pi^2 \text{ and } 	M_-^\text{R}=\Pi^1 .
\end{equation}
This illustrates that, whenever we use the NIMP to measure $\rep C$ and $O_1$ satisfies \eqref{e:spectrum}, the ancilla-based measurement is equivalent to a projective measurement of the target in the eigenbasis of $O_1$. In this special case, $\rep \matel{\psi}{O_1(t_1)O_2(t_2)}{\psi}$ can be measured by projectively probing the target at $t_1$ in the eigenbasis of $O_1$, and then at $t_2$ in the eigenbasis of $O_2$.

Evidently, spin-$1/2$ correlations of the form $\matel{\psi}{S_i^a(t_1)O_2(t_2)}{\psi}$ are a special case in which noninvasive measurements can be avoided: The imaginary part can be measured by performing a local rotation $\exp(-i\theta S_i^a)$ of the target state at $t_1$, followed by a projective measurement of $O_2$ at $t_2$. The real part can be obtained by projectively measuring $S_i^a$ at $t_1$, followed by a projective measurement of $O_2$ at $t_2$. For a detailed discussion of ancilla-free measurement protocols for these special spin-$1/2$ correlations we refer the reader to Secs.~VII and VIII as well as Appendix~B of \cite{Uhrich_etal}.

\section{Conclusions}
In this paper we have elaborated on ideas, first introduced in Ref.~\cite{Uhrich_etal}, on how to reduce (or even entirely avoid) the effect of measurement backaction when measuring dynamic correlation functions in quantum spin systems. We have generalised the noninvasive measurement protocol of Ref.~\cite{Uhrich_etal} in various ways, showing how to obtain correlations between arbitrary multi-site observables and relaxing constraints on the spin quantum number of the ancilla.

We have introduced an alternative approach in which two ancillas are simultaneously coupled to the target system at time $t_1$ and subsequently measured, with the goal of extracting both $\rep C$ and $\imp C$ from the same set of measurements. From an experimental viewpoint this translates into fewer required resources such as time and manpower, which may make the simultaneous coupling version of the NIMP advantageous in some circumstances (assuming that coupling two ancillas to the target is not much more difficult than coupling only one).

By deriving the POVM corresponding to the ancilla-based noninvasive measurements we were able to gain a better understanding of their effect on the target. We found that the effect of the NIMP when measuring $\imp C$ is to unitarily evolve the target state for a time $\lambda$, with the generating Hamiltonian given by the observable $O_1$ which is to be correlated at $t_1$. As a result we were able to show in Eq.~\eqref{e:imRot} that the imaginary part of any dynamic correlation can be measured without an ancilla by performing a suitable unitary evolution and in the limit of $|\lambda|\ll 1$.
For $\rep C$ the Kraus operators \eqref{e:krausRe} reveal that the noninvasive measurement causes a simultaneous projection of the target state onto all eigenspaces of $O_1$. We derived a spectral condition \eqref{e:spectrum} for the correlated observable $O_1$ under which $M_\pm^\text{R}$ reduces to the individual projection operators corresponding to a projective measurement of $O_1$ \eqref{e:krausReSpecial}. Correlations in which $O_1$ has the required spectrum can thus be measured without an ancilla, purely by means of projective measurements. The POVM formalism has thus allowed us to motivate the ancilla-free measurement protocols for correlations $\matel{\psi}{S_i^a(t_1)O_2(t_2)}{\psi}$, which were previously presented in \cite{Uhrich_etal}.
%
%
\begin{acknowledgement}
M.K.\ acknowledges financial support from the National Research Foundation of South Africa via the Incentive Funding and the Competitive Programme for Rated Researchers. P.U.\ acknowledges financial support from the National Research Foundation of South Africa, Stellenbosch University as well as the Sam Cohen Trust.

\end{acknowledgement}
\appendix{
\section{Ancilla states for simultaneous noninvasive measurements}
\label{app:proofB2}
For the simultaneous noninvasive measurement protocol of Sec.~\ref{sec:simCouple}---in which both ancillas are measured in the $S^z$ eigenbasis---we have the following conditions on the initial ancilla states,
\begin{equation}\label{e:ancCond}
	\expval{S^z}_{\phi_1}=0,\quad\expval{S^z}_{\phi_2}=0,\quad\text{and $\expval{B_2}_{\phi_2}=0$ with $B_2=S^y$},
\end{equation}
where the spin quantum number $\zeta\in\NN/2$ of both ancillas can be arbitrary. The $S^z$ expectation values are zero if both ancilla initial states satisfy \eqref{e:phi}, and it is sufficient to assume both these states to be the same, $\ket{\phi_1}=\ket{\phi_2}=\ket{\phi}$. It remains to ensure that the last condition of \eqref{e:ancCond}, $\expval{S^y}_{\phi_2}=0$, is satisfied. To do so we write $S^y$ in terms of the spin-$\zeta$ ladder operators
\begin{equation}\label{e:ladder}
	S^y=(-i/2)(S^+-S^-)
\end{equation}
with
\begin{equation}
    S^{\pm} = \sum_{m,m'=-\zeta}^\zeta c_{\pm}(\zeta,m)\ket{m'}\!\bra{m}\delta_{m',m\pm1},
\end{equation}
where $c_\pm(\zeta,m) = \sqrt{\zeta(\zeta+1)-m(m\pm1)}$. We then have
\begin{equation}
	\begin{split}
	\expval{S^y}_{\phi}=&(-i/2)\sum_{m,m'=-\zeta}^\zeta r_m r_{m'} e^{i(\theta_{m'}-\theta_{m})}\left[ \matel{m}{S^+}{m'}-\matel{m}{S^-}{m'}\right]\\
	\propto& \sum_{m,m'=-\zeta}^\zeta r_m r_{m'} e^{i(\theta_{m'}-\theta_{m})}\left[ c_+(\zeta,m')\delta_{m,m'+1}-c_-(\zeta,m')\delta_{m,m'-1}\right] \\
	=& \sum_{m=-\zeta}^\zeta r_{m}\left[ r_{m-1} e^{i(\theta_{m-1}-\theta_{m})} c_+(\zeta,m-1)- r_{m+1} e^{i(\theta_{m+1}-\theta_{m})}c_-(\zeta,m+1) \right] \\
	=& \sum_{m=-\zeta}^{\zeta-1} r_{m+1} r_{m} \left[e^{i(\theta_{m}-\theta_{m+1})} c_+(\zeta,m)- e^{i(\theta_{m+1}-\theta_{m})}c_-(\zeta,m+1) \right] \\
	\propto& \sum_{m=-\zeta}^{\zeta-1} r_{m+1} r_{m}  c_+(\zeta,m) \impart{e^{i(\theta_{m}-\theta_{m+1})} }.
	\end{split}
\end{equation}
In the fourth line we have used $c_-(\zeta,\zeta+1)=c_+(\zeta,-\zeta-1)=0$, and in the last line we have used that $c_-(\zeta,m+1)=c_+(\zeta,m)$. For the summation in the last line to be zero, the simplest option is to insist that $\impart{e^{i(\theta_{m}-\theta_{m+1})}}=0$ for all $m\in\{-\zeta,\ldots,\zeta\}$, which leads to the condition $\theta_{m}=\theta_{m+1}+k_m\pi$ for any integer $k_m$. This implies that, up to a global phase, we must use initial ancilla states
\begin{equation}
	\ket{\phi} = \sum_{m=-\zeta}^\zeta e^{ik_m\pi}r_m \ket{m}=\sum_{m=-\zeta}^\zeta (-1)^{k_m}r_m \ket{m} \quad\text{with $r_m=r_{-m}$ for all $m$}.
\end{equation}
}

\bibliographystyle{unsrt}
\bibliography{GranBib}
%
\end{document}